\documentclass[%
 reprint,
 amsmath,amssymb,
 aps,
]{revtex4-2}

\usepackage{graphicx}
\usepackage{dcolumn}
\usepackage{bm}

\def\s{{\mathbf{s}}}

\def\refe#1{Eq.~(\ref{#1})}

\begin{document}

\title{Crossing Symmetric Dispersion Relations without Spurious Singularities}

\author{Chaoming Song}%
 \email{c.song@miami.edu}
\affiliation{%
Department of Physics, University of Miami, Coral Gables FL, 33146 USA.
}%

\date{\today}

\begin{abstract}
Recently, there has been renewed interest in a crossing-symmetric dispersion relation from the 1970s due to its implications for both regular quantum field theory and conformal field theory. However, this dispersion relation introduces nonlocal spurious singularities and requires additional locality constraints for their removal, a process that presents considerable technical challenges. In this Letter, we address this issue by deriving a new crossing-symmetric dispersion relation that is free of spurious singularities, resulting in  a compact form of the contact terms in crossing-symmetric blocks. Our results establish a solid foundation for the Polyakov bootstrap in conformal field theories and the crossing-symmetry S-matrix bootstrap in quantum field theories.
\end{abstract}

\maketitle

{\bf Introduction:} 
The recent revival in crossing-symmetric dispersion relations~\cite{auberson1972rigorous,mahoux1974physical} has sparked considerable interest in both quantum field theory (QFT)~\cite{sinha2021crossing} and conformal field theory (CFT)~\cite{gopakumar2021crossing,bissi2022selected}. In contrast to traditional $t$-fixed dispersion relations, which display symmetry in only two channels~\cite{mandelstam1958determination,nussenzveig1972causality}, crossing-symmetric dispersion relations impose no additional constraints and are in perfect accord with Feynman diagram expansions. Within the CFT domain, four-point correlation functions must adhere to crossing symmetry constraints. Numerical bootstrap typically enforces this crossing symmetry on the conformal block expansion. Alternately, Polyakov introduced a conformal bootstrap using crossing-symmetric blocks~\cite{polyakov1974nonhamiltonian}, an approach that has recently proven effective in Mellin space~\cite{sen2016critical,gopakumar2017conformal,gopakumar2017mellin}. This method employs a basis connected to exchange Witten diagrams, although contact terms remain undetermined~\cite{dey2017mellin,dey2018simplifying}. Resolving these terms continues to pose a considerable challenge~\cite{gopakumar2018polyakov,mazavc2019analytic,mazavc2019crossing,ferrero2020crossing,penedones2020nonperturbative,carmi2021applications,caron2021dispersive,sleight2020unique}.

Gopakumar et al.\cite{gopakumar2021crossing} recently observe that these contact term ambiguities are fully determined using a crossing-symmetric dispersion relation, initially developed by Auberson and Khuri (AK)~\cite{auberson1972rigorous} and later revisited by Sinha and Zahed~\cite{sinha2021crossing}. However, the AK dispersion relation presents spurious singularities that violate locality. Therefore, additional locality constraints are manually imposed to remove these unphysical terms. In theory, after removing these singularities, crossing-symmetric dispersion relations allow for a Feynman/Witten diagram expansion and entirely fix the contact terms. In line with this approach, a closed form of the contact terms has been proposed~\cite{chowdhury2022locality}. Nevertheless, the complexity of analyzing singularities restricts its practical application to lower spins, thereby complicating the implementation of the Polyakov bootstrap.

In this paper, we propose a new dispersion relation that manifests both crossing symmetry and locality. We discover a novel approach to directly remove nonlocal singularities, resulting in a closed form of the singularity-free dispersion relation. Consequently, we present the Feynman/Witten expansion of crossing-symmetric blocks, providing explicit determinations of all contact terms. Furthermore, we develop the full dispersion relation without assuming crossing-symmetric amplitudes, enabling the application of our findings to a wide range of problems. For instance, our work establishes a solid foundation for the Polyakov bootstrap, where the only remaining non-trivial constraint is the Polyakov condition~\cite{polyakov1974nonhamiltonian,gopakumar2017conformal}. Moreover, our approach yields a novel functional sum rule for the crossing-symmetric bootstrap, eliminating the need for power series expansions.

{\bf Singularity-free dispersion relation:} 
We begin with the shifted Mandelstam variables $s_1 = s-\mu/3$, $s_2 = t-\mu/3$, and $s_3 = u-\mu/3$ satisfying the constraint $s_1 + s_2 + s_3 = 0$, where $s$, $t$, and $u$ are the usual Mandelstam variables. For regular QFT, we have $\mu = 4m^2$, while for CFT, we have $\mu = 2\Delta_\phi$. We consider hypersurfaces $(s_1-a)(s_2-a)(s_3-a) = -a^3$, and rewrite $s_k(z,a) = a - a(z-z_k)^3/(z^3-1)$, where $z_k$ are cube roots of unity~\cite{auberson1972rigorous}. Note that we can express $a = y/x$, where $x \equiv -( s_1s_2 + s_2 s_3 + s_3 s_1)$ and $y \equiv - s_1 s_2 s_3$. Instead of a dispersion relation in $s$ for fixed $t$, we can write down a twice subtracted dispersion relation in the variable $z$, for fixed $a$. The full crossing-symmetric dispersion relation is quite involved, and we refer the readers to Ref.~\cite{auberson1972rigorous} for more details. A full singularity-free dispersion relation is set to be proposed in a subsequent section. 

Our discussion below primarily focuses on the completely crossing-symmetric scattering amplitudes, such as pion-pion scattering in QFT or the Mellin amplitude for a four-point correlation of identical scalars in CFT \cite{mack2009d,penedones2011writing}. For a crossing-symmetric amplitude $\mathcal{M}^{(s)}$, the dispersion relation simplifies dramatically in terms of $\s \equiv \{s_1, s_2, s_3\}$, as
\begin{equation}
\begin{split}
    \mathcal{M}^{(s)}(\s) = \alpha_0 + \frac{1}{\pi}  \int&   \frac{d\sigma}{\sigma}  \mathcal{A}\left(\sigma, s_\pm\left(\sigma,  \frac {a}{\sigma-a} \right)\right)  \\ 
    &   \times H(\sigma;\s),
\end{split}
\label{eq:AK}
\end{equation}
where $\mathcal{A}(s_1,s_2)$ is the s-channel discontinuity, symmetric under the exchange of $t$ and $u$ channels, i.e., $\mathcal{A}(s_1,s_2) = \mathcal{A}(s_1,s_3)$. The constant $\alpha_0 \equiv \mathcal{M}^{(s)}(0,0)$, and the functions $H(\sigma;\s)$ and $s_\pm(\sigma,\eta)$ are defined as:
\begin{eqnarray}
H(\sigma;\s) &\equiv& \frac{s_1}{\sigma - s_1} + \frac{s_2}{\sigma - s_2} + \frac{s_3}{\sigma - s_3},  \nonumber \\
s_\pm(\sigma,\eta) &\equiv& \sigma\frac{-1 \pm \sqrt{1+4 \eta}}{2},  \nonumber    
\end{eqnarray}
where $s_+s_- = -\sigma^2 \eta $, and $s_+ + s_- = - \sigma$. Setting $\eta = a/(\sigma-a)$ and $s_1 = \sigma$ solves $s_2 = s_\pm$ and $s_3 = s_\mp$ from the definition above. Note that $\mathcal{A}(\sigma, s_+) = \mathcal{A}(\sigma, s_-)$, and thus the validity of \refe{eq:AK} is independent of the choice of $s_+$ or $s_-$. 

Equation~(\ref{eq:AK}) is manifestly crossing symmetric, allowing the scattering amplitude 
\begin{equation}\label{eq:MW}
    \mathcal{M}^{(s)}(\s) = \sum_{p,q} \mathcal{M}^{(s)}_{p,q} x^p y^q, 
\end{equation}
to be expanded in terms of crossing-symmetric variables $x$ and $y$. However, the AK dispersion relation~(\ref{eq:AK}) involves the variable $a$ and, therefore, leads to negative powers of $x$ in the expansion~(\ref{eq:MW}). These spurious singularities are known to violate locality \cite{sinha2021crossing}. To obtain the physical scattering amplitude, additional locality constraints must be imposed to enforce the vanishing of these non-physical terms in \refe{eq:MW}. Formally, a singularity-free dispersion relation requires computing the regular part
\begin{align}\label{eq:R}
   R \equiv  \mathcal{R}\left\{\mathcal{A}\left(\sigma, s_\pm\left(\sigma,  \frac {a}{\sigma-a} \right)\right)  H(\sigma;\s) \right\},
\end{align}
where $\mathcal{R} \{\ldots\}$ denotes a formal regularization with the negative power of $x$ terms being removed.

To obtain a closed form of the regular part $R$, we first rewrite $H(\sigma,\s) = (2\sigma^3 -y)H_0(\sigma,\s)-2$, where
\begin{equation}
    H_0(\sigma,\s) \equiv  \frac{1}{(\sigma-s_1)(\sigma-s_2)(\sigma-s_3)} = \frac{1}{\sigma^3 + y - \sigma x }, \nonumber
\end{equation}
corresponds to the poles. Notice that multiplying the factor $a$ with a regular function $f(x,y)$,
\begin{equation}
    \hat a f(x,y) \equiv \mathcal{R}\{a f(x,y)\}\nonumber
\end{equation}
acts as a lowering operator $\hat a |n\rangle = y |n-1\rangle$, with $\hat a|0\rangle = 0$, where $|n\rangle \equiv x^n$ denotes the $n$-th power of $x$. Specifically, we obtain
\begin{align}
    \hat a^n H_0 & = \frac{1}{\sigma^3+y} \sum_{m=0}^\infty  \left( \frac{\sigma }{\sigma^3+y}\right)^m \hat a^n x^m \nonumber\\
    &=  \frac{1}{\sigma^3+y} \sum_{m=n}^\infty  \left( \frac{\sigma }{\sigma^3+y}\right)^m  y^n x^{m-n} \nonumber\\
    &= \left( \frac{\sigma y}{\sigma^3+y}\right)^n H_0, \nonumber
\end{align}
which suggests that 
\begin{equation}\label{eq:F}
    F(\hat a,y) H_0(\sigma,\s) = F\left(\frac{\sigma y}{\sigma^3+y}, y\right) H_0(\sigma,\s) 
\end{equation}
for any function $F(a,y)$ admitting a Taylor expansion in terms of $a$. Substituting \refe{eq:F} into \refe{eq:R} and noting $F(\hat a,y) f(y) = F(0,y) f(y)$ lead to
\begin{equation}
R =\mathcal{A}\left(\sigma, s_\pm\left(\sigma,  y/\sigma^3 \right)\right)  (2\sigma^3 -y)H_0(\sigma,\s) - 2 \mathcal{A}\left(\sigma, s_\pm\left(\sigma, 0 \right)\right). \nonumber
\end{equation}
Therefore, we obtain the singularity-free (SF) dispersion relation 
\begin{widetext}
\begin{equation}\label{eq:our}
\mathcal{M}^{(s)}(\s) = \alpha_0 + \frac{1}{\pi} \int \frac{d\sigma}{\sigma}  \left(   \frac{(2\sigma^3 +s_1s_2s_3)\mathcal{A}\left(\sigma, s_\pm\left(\sigma, -s_1s_2s_3/\sigma^3\right)\right)}{(\sigma-s_1)(\sigma-s_2)(\sigma-s_3)}  - 2 \mathcal{A}(\sigma,0) \right),
\end{equation}
\end{widetext}
where the locality constraints are automatically satisfied, as we will show explicitly in the next section.

{\bf Block expansion and contact terms:}
To facilitate the analysis of the s-channel discontinuity, it is common practice to expand it in terms of the partial waves with {\it even} spins, as 
\begin{equation}
 \mathcal{A}(s_1, s_2) = \sum_\ell \int d \lambda f_\ell(\sigma,\lambda) Q_{\lambda,\ell}(s_1, s_2),\nonumber
\end{equation}
where the partial wave $Q_{\lambda,\ell}(s_1, s_2) = Q_{\lambda,\ell}(s_1, s_3)$ is a symmetric polynomial of order $\ell$ that is invariant under the exchange of the $ut$ channels, and the spectrum $f_\ell(\sigma,\lambda)$ encodes scattering data. For QFT, we express $Q_{0,\ell}(s_1, s_2) \equiv (s_1-2\mu/3)^{\ell}C_\ell^{(\frac{d-3}{2})}\left(\frac{s2-s3}{s_1-2\mu/3}\right)$ in terms of Gegenbauer polynomials, and $f_\ell(\sigma,\lambda) = (\sigma-2\mu/3)^{-\ell}\Phi(\sigma)(2\ell+2\alpha)\alpha_\ell(\sigma)\delta(\lambda)$ with $\Phi(\sigma) \equiv \Psi((d-3)/2)(\sigma+\mu)^{1/2}/(\sigma-2\mu/3)^{(d-3)/2}$ with a real number $\Psi((d-3)/2)$. For CFT, we express $Q_{\Delta,\ell}(\s) = P_{\Delta-d/2,\ell}(s_1+2\Delta_\phi/3, s_2-\Delta_\phi/3)$ in terms of Mack polynomials, and $f_\ell(\sigma,\lambda) \equiv \sum_{\Delta,k} C_{\Delta,\ell} N_{\Delta,k} \delta(\sigma - s_k) \delta(\lambda-\Delta)$ encodes the operator product expansion (OPE) data.

The scattering amplitude can also be expressed as
\begin{equation}
\mathcal{M}^{(s)}(\s) = \alpha_0 + \frac{1}{\pi}\sum_{\ell=0}^\infty \int  d\sigma d\lambda f_\ell(\sigma,\lambda) M_{\lambda,\ell}(\sigma; \s), \nonumber
\end{equation}
where $ M_{\lambda,\ell}(\sigma; \s)$ are scattering blocks. Comparing AK dispersion relation~(\ref{eq:AK}), we obtain the Dyson block~\cite{sinha2021crossing},
\begin{equation}\label{eq:DB}
M^{(D)}_{\lambda,\ell}  = \frac{1}{\sigma} Q_{\lambda,\ell}\left(\sigma, s_\pm\left(\sigma,  \frac {a}{\sigma-a} \right)\right)H(\sigma;\s),
\end{equation}
which contains spurious singularities. By contrast,
our dispersion relation~(\ref{eq:our}) leads to the singularity-free block
\begin{equation}\label{eq:FB}
M^{(SF)}_{\lambda,\ell} =\frac{(2\sigma^3 -y)Q_{\lambda,\ell}(\sigma, s_\pm(\sigma, y/\sigma^3))}{\sigma(\sigma-s_1)(\sigma-s_2)(\sigma-s_3)}  -\frac{2}{\sigma}Q_{\lambda,\ell}(\sigma, 0). 
\end{equation}

To show explicitly SF block $M^{(SF)}_{\lambda,\ell}$ removes spurious singularities present in the Dyson block $M^{(D)}_{\lambda,\ell}$, we take the QFT case as an example. We start with the Gegenbauer polynomials $C_\ell^{(\frac{d-3}{2})}(\sqrt{\xi})$  where $\xi = (s_+(\sigma,\eta) - s_-(\sigma,\eta))^2/(\sigma_1-2\mu/3)^2 = \xi_0 (1+ 4\eta)$, where $\xi_0 \equiv \sigma^2/(\sigma-2\mu/3)^2$. We set $\eta = a/(\sigma-a)$ and expand the Gegenbauer polynomials around $\xi_0$, giving us \cite{auberson1972rigorous,sinha2021crossing}
\begin{equation}
    M^{(D)}_{\lambda,\ell}  = \frac{1}{\sigma} \sum_{n,m=0}^\infty  \mathcal{B}_{n,m}^{(\ell)} x^n (y/x)^m , \nonumber 
\end{equation}
where $ p_\ell^{(k)} \equiv \partial^k_\xi C^{(d-3)/2}(\sqrt\xi_0)$, and 
\begin{equation}
\mathcal{B}_{n,m}^{(\ell)} = \sum_{k=0}^m \frac{ p_\ell^{(k)}(4\xi_0)^k (3k-m-2n)(-n)_m}{\pi \sigma^{2n+m}k!(m-k)!(-n)_{k+1}}.\nonumber
\end{equation}

Similarly, expanding the Gegenbauer polynomials around $\xi_0$ with $\eta = y/\sigma^3$ leads to
\begin{equation}
    M^{(SF)}_{\lambda,\ell}  = \frac{1}{\sigma} \sum_{n,m=0}^\infty  \mathcal{C}_{n,m}^{(\ell)} x^n y^m,  \nonumber
\end{equation}
where
\begin{align}
\mathcal{C}_{n,m}^{(\ell)} &= \sum_{k=0}^m \frac{  p_\ell^{(k)}(4\xi_0)^k  (-1)^{m-k} (2n+3(m-k))}{\pi \sigma^{2n+3m}n!k!(m-k)!}\nonumber\\
&\times (n+m-k-1)!.\nonumber
\end{align}
It is easy to verify that $\mathcal{C}_{n,m}^{(\ell)} = \mathcal{B}_{n+m,m}^{(\ell)}$ for $n,m \geq 0$, indicating that the regular part of the Dyson blocks matches with the SF blocks, as expected. However, the Dyson blocks have spurious singularities with a negative power of $x$ when $n < m$, which is absent in our SF blocks. A similar deviation can be obtained for general partial waves $Q_{\lambda,\ell}$.

The singularity-free (SF) block provides a block expansion for the amplitude that directly relates to the usual Feynman and Witten diagrammatic expansions for QFT and CFT, respectively. To see this, we will show below that the SF block can be written as a summation of exchange and contact terms, as follows:
\begin{equation}\label{eq:expansion}
M^{(SF)}_{\lambda,\ell}(\sigma;\s) = \sum_{i=1}^3 M_{\lambda,\ell}^{(i)}(\sigma;\s) +  M_{\lambda,\ell}^{(c)}(\sigma;\s), 
\end{equation}
where the exchange term of channel $i$ is given by 
\begin{equation}
M_{\lambda,\ell}^{(i)}(\sigma;\s) = Q_{\lambda,\ell}(s_{i},s_{i+1})\left(\frac{1}{\sigma - s_{i}}-\frac{1}{\sigma}\right),\nonumber
\end{equation}
for $i = 1,2,3$ with the cyclic condition $i+1 = 1$ for $i=3$. 
The contact terms $ M_{\lambda,\ell}^{(c)}(\sigma;\s)$ involve polynomials of $s_i$'s, whose explicit form is previously only known for few lower order terms. 

We substitute \refe{eq:FB} into \refe{eq:expansion}, obtaining
\begin{equation}\label{eq:Mc}
    M_{\lambda,\ell}^{(c)}(\sigma;\s) = \frac{1}{\sigma} \sum_{i=1}^3 \frac{s_i \Delta Q_{\lambda,\ell}^{(i)}}{(\sigma-s_i)} + \frac{2}{\sigma}\Delta Q_{\lambda,\ell}^{(0)},
\end{equation}
where 
\begin{align}
   &\Delta Q_{\lambda,\ell}^{(i)} \equiv Q_{\lambda,\ell}(\sigma;s_\pm(\sigma, y/\sigma^3) ) - Q_{\lambda,\ell}(s_i; s_\pm(s_i, y/s_i^3) ), \nonumber \\ 
   &\Delta Q_{\lambda,\ell}^{(0)}\equiv Q_{\lambda,\ell}(\sigma, s_\pm(\sigma, y/\sigma^3)) -  Q_{\lambda,\ell}(\sigma, 0), \nonumber
\end{align}
are polynomials. To show the the contact term $M_{\lambda,\ell}^{(c)}$ are also polynomials,  we notice that the symmetry of ut channels allows us to expand $Q_{\lambda,\ell}(s_1;s_2 )= \sum_{n+2m \leq l} q_{nm} s_1^n (s_2 s_3)^{m}$, 
which implies
\begin{equation}
Q_{\lambda,\ell}(\sigma; s_\pm(\sigma, y/\sigma)) = 
\sum_{n+2m \leq l} q_{nm} \sigma^n (s_1/\sigma)^{m} (s_2 s_3)^{m}. \nonumber
\end{equation}
Thus, 
\begin{equation}
\Delta Q_{\lambda,\ell}^{(i)} 
= \sum_{n,m} q_{nm} \sigma^{n} \left(  (s_i/\sigma)^m   -  (s_i/\sigma)^n    \right)  (s_{i+1} s_{i+2})^m, \nonumber
\end{equation}
where the term $ (s_i/\sigma)^m   -  (s_i/\sigma)^n $ must divide $s_i/\sigma-1$ and thus cancel the poles in \refe{eq:Mc}. More explicitly, we find that 
\begin{equation}
\Delta Q_{\lambda,\ell}^{(i)} 
= (s_i-\sigma) \sum_{n,m} P_{n,m}(\sigma) s_i^n (s_{i+1}s_{i+2})^m,  
\nonumber\end{equation}
where
\begin{equation}
     P_{n,m}(\sigma) =
     \begin{cases}
         \sum_{k=0}^n q_{km} \sigma^{k-n-1}, &0\leq n < \min(m,\ell-2m+1) \\
         \sum_{k=0}^{\ell-2m} q_{km} \sigma^{k-n-1}, & \ell-2m \leq n \leq m-1, \\
         -\sum_{k={n+1}}^{\ell-2m} q_{km} \sigma^{k-n-1}, & m \leq n \leq \ell-2m-1.\nonumber
     \end{cases}
\end{equation}
Substituting into \refe{eq:Mc}, we obtain the contact term
\begin{eqnarray}
M_{\lambda,\ell}^{(c)} = \frac{2}{\sigma}\left (Q_{\lambda, \ell} (\sigma, s_\pm(\sigma,y/\sigma^3)) - Q_{\lambda, \ell} (\sigma, 0) \right) \nonumber \\- \frac{1}{\sigma}  \sum_{n,m} P_{n,m}(\sigma)\left(\sum_{i=1}^3s_{i}^{n+1}(s_{i+1}s_{i+2})^m\right),
\end{eqnarray}
which are manifestly crossing-symmetric polynomials.  Note that the summation over indices $n$ and $m$ is across all non-zero $P(n,m)$ terms, i.e., $0 \leq m \leq \ell/2$ and $n+2m \leq 3\ell/2-1$. 

{\bf Singular block and sum rules:} 
Since the SF block corresponds to the regular part of the Dyson block, we can decompose 
\begin{equation}
    M^{(D)}_{\lambda,\ell}(\sigma; \s) = M_{\lambda,\ell}^{(SF)}(\sigma; \s) + M_{\lambda,\ell}^{(S)}(\sigma; \s), \nonumber
\end{equation}
where $M_{\lambda,\ell}^{(S)}(\sigma; \s)$ refers to the corresponding singular part, given by
\begin{align}
M^{(S)}_{\lambda,\ell} &= \frac{Q_{\lambda,\ell}(\sigma,s_\pm(\sigma,y/\sigma^3)-Q_{\lambda,\ell}(\sigma,s_\pm(\sigma,a/(\sigma-a)))}{y/\sigma^3-a/(\sigma-a)} \nonumber \\
 &  \times \frac{a(2\sigma x-3y)}{\sigma^4(\sigma-a)} - \frac{2}{\sigma} (Q_{\lambda,\ell}(\sigma,s_\pm(\sigma,y/\sigma^3) - Q_{\lambda,\ell}(\sigma,0)). \nonumber
\end{align}
Note since $Q_{\lambda,\ell}(\sigma,s_\pm(\sigma,\eta))$ is polynomial of $\eta$, the term in the first line is the difference operator acting on $Q_{\lambda,\ell}$ between $\eta = y/\sigma^3$ and $\eta = a/(\sigma-a)$, and thus is also a polynomial of these two terms. Therefore, the first term involves positive powers of $y/x$ except for the zeroth-order term $(y/x) 2\sigma x$, which cancels the last term in the above equation. Consequently, only terms with negative powers of $x$ remain in $M^{(S)}_{\lambda,\ell}$, as expected.
 
Since both Dyson and SF blocks lead to the same amplitude, the contribution from the singular part needs to be canceled:
\begin{equation}\label{eq:sum}
\sum_{\ell} \int  d\sigma d\lambda f_\ell(\sigma,\lambda) M^{(S)}_{\lambda,\ell}(\sigma; \s) = 0,
\end{equation}
which imposes a constraint on the spectrum $f_\ell(\sigma,\lambda)$. For instance, for QFT, Equation~(\ref{eq:sum}) requires the cancellation of power series contributions of $\mathcal{B}^{(\ell)}_{n,m} x^{n-m} y^m$ with negative powers of $x$, i.e., $n < m$, generalizing the Froissart bound \cite{sinha2021crossing,tolley2021new}. For CFT, it appears to connect to the conformal dispersive sum rules \cite{carmi2021applications,gopakumar2021crossing}. Unlike previous approaches, Eq.~(\ref{eq:sum}) provides a single functional sum rule without involving series expansion.

{\bf Full dispersion relation:}
Our approach extends to general scattering amplitudes without assuming the complete crossing symmetry. The corresponding full dispersion relation should link the scattering amplitude $\mathcal{M}(\s)$ to $s$, $u$, and $t$-channel discontinuities, denoted as $\mathcal{A}_i(\s)$ for $i=1,2,3$. Furthermore, $\mathcal{M}(\s)$ is not merely a function of $x$ and $y$, but also of a linear combination of $s_i$. In addition, an antisymmetric part exists \cite{kaviraj2022crossing}, characterized in terms of $w = - (s_1-s_2)(s_2-s_3)(s_3-s_1)$~. Note that the algebraic curve $w^2 = 4x^3 - 27 y^2$ suggests that any power of $w$ higher than first order will be absorbed in a combination of $x$ and $y$. Using an approach similar to the one presented above, we derive the full SF dispersion relation:
\begin{widetext}
\begin{align}\label{eq:our1}
\mathcal{M}(\s) = \alpha_0 +\sum_{i=1}^3 \alpha_i s_i + \frac{1}{2\pi} \sum_{i=1}^3  \int \frac{d\sigma}{\sigma}  \frac{ K_i^+(\s,\sigma) \mathcal{A}_i\left(\sigma, \tilde s_+\right) + 
K_i^-(\s,\sigma) \mathcal{A}_i\left(\sigma, \tilde s_-\right)
}{(\sigma-s_1)(\sigma-s_2)(\sigma-s_3)} 
 \\ 
 -  K_i^{0+}(\s,\sigma) \mathcal{A}_i\left(\sigma, s_+(\sigma,0)\right) -  K_i^{0-}(\s,\sigma) \mathcal{A}_i\left(\sigma, s_-(\sigma,0)\right), \nonumber
\end{align}
\end{widetext}
where $\tilde s_\pm \equiv s_\pm(\sigma, y/\sigma^3)$, $K_i^{0\pm}(\s,\sigma) =\frac{2}{3}+\frac{s_i}{\sigma}   \pm \frac{s_{i+1}-s_{i+2}}{3\sigma}$, and
\begin{align}
   K_i^\pm(\s,\sigma) &=  \left(  (2\sigma^3 -y) \pm \frac{\sigma w}{\tilde s_+ - \tilde s_-} \right) \left(\frac{1}{3}+\frac{\sigma^2 s_i }{2(y+\sigma^3)}  \right) \nonumber \\
   &+ \left( \sigma^2 w  \pm \frac{(2\sigma^3-y)(4y+\sigma^3)}{\tilde s_+ - \tilde s_-}  \right) \frac{s_{i+1}-s_{i+2}}{6(y+\sigma^3)}. \nonumber 
\end{align}
The constants $\alpha_i$ correspond to the first-order coefficient of $s_i$, with only two being free, enabling us to impose $\sum_{i=1}^3 \alpha_i = 0$. The corresponding SF blocks can be found consequently. The full dispersion relation~(\ref{eq:our1}) is substantially more involved. 

While the full dispersion relation~(\ref{eq:our1}) is considerably more involved, it simplifies remarkably for the crossing symmetric and symmetric cases. In the former scenario, the discontinuities across all channels are identical, i.e., $\mathcal{A}_i(\sigma, \tilde s_\pm) = \mathcal{A}(\sigma, \tilde s_\pm)$. Equation~(\ref{eq:our1}) reduces to \refe{eq:our} since all terms cancel after summation except for $(2\sigma^3-y)/3$. Likewise, for the crossing antisymmetric case~\cite{kaviraj2022crossing}, Equation~(\ref{eq:our1}) simplifies to 
\begin{equation}\label{eq:ourAS}
\mathcal{M}^{(as)}(\s) = \frac{w}{\pi} \int \frac{ d\sigma}{(\sigma-s_1)(\sigma-s_2)(\sigma-s_3)}     \frac{\mathcal{A}(\sigma, \tilde s_+)}{\tilde s_+-\tilde s_-}, \nonumber
\end{equation}
which provides the crossing antisymmetric dispersion relation. It is noteworthy that in this case $\mathcal{A}(\sigma, \tilde s_+) = -\mathcal{A}(\sigma, \tilde s_-)$, thus $\frac{\mathcal{A}(\sigma, \tilde s_+)}{\tilde s_+-\tilde s_-} = \frac{1}{2} \frac{\mathcal{A}(\sigma, \tilde s_+)-\mathcal{A}(\sigma, \tilde s_-)}{\tilde s_+-\tilde s_-}$ is a polynomial in terms of $\tilde s_+$ and $\tilde s_-$, as expected for {\it odd} spin contributions. 
 
{\bf Discussion:}
The singularity-free, crossing-symmetric dispersion relation approach introduced in this paper addresses a long-standing challenge in the nonperturbative exploration of quantum field theories. The proposed crossing-symmetric blocks seamlessly link to Feynman/Witten diagrams, with contact terms being explicitly determined. The null contribution of the singular block leads to a simplified functional sum rule, enhancing existing methods. Furthermore, the full singularity-free dispersion relation lays the groundwork for the Polyakov bootstrap beyond identity operators. This approach also provides remarkable opportunities for numerical S-matrix bootstrap within a broader setup. Undoubtedly, our advancements establish a robust foundation for crossing-symmetric bootstrap applicable to both QFTs and CFTs.

\begin{acknowledgments}
We express our gratitude to Aninda Sinha and Ahmadullah Zahed for their invaluable discussions and constructive feedback on the manuscript.
\end{acknowledgments}

\end{document}